\documentclass[preprint]{aastex62} 
\usepackage{mathtools}
\usepackage{booktabs}
\usepackage{multirow}

\newcommand{\psm}{\texttt{PSM}}
\newcommand{\ipsm}{\texttt{iPSM}}

\graphicspath{{./}{figures/}}

\submitjournal{ApJL}

\shorttitle{XSM stuff}
\shortauthors{Upendran et al.}

\begin{document}

\title{Nanoflare Heating of the Solar Corona Observed in X-rays}

\correspondingauthor{Vishal Upendran}
\email{uvishal@iucaa.in}

\author[0000-0002-9253-6093]{Vishal Upendran}
\author[0000-0003-1689-6254]{Durgesh Tripathi}
\affiliation{Inter University Centre for Astronomy and Astrophysics, Pune, India - 411007}

\author[0000-0003-3431-6110]{N.P.S. Mithun}
\affiliation{Physical Research Laboratory, Navrangpura, Ahmedabad, Gujarat-380 009, India}
\author[0000-0002-2050-0913]{Santosh Vadawale}
\author[0000-0003-1693-453X]{Anil Bhardwaj}
\affiliation{Physical Research Laboratory, Navrangpura, Ahmedabad, Gujarat-380 009, India}
\begin{abstract}

The existence of the million-degree corona above the cooler photosphere is an unsolved problem in astrophysics. Detailed study of quiescent corona that exists regardless of the phase of the solar cycle may provide fruitful hints towards resolving this conundrum. However, the properties of heating mechanisms can be obtained only statistically in these regions due to their unresolved nature. Here, we develop a two-step inversion scheme based on the machine learning scheme of \cite{VDModel} for the empirical impulsive heating model of \cite{PSModel}, and apply it to disk integrated flux measurements of the quiet corona as measured by the X-ray solar monitor (XSM) onboard Chandrayaan{--}2. We use data in three energy passbands, \textit{viz.}, 1{--}1.3~keV, 1.3{--}2.3~keV, and 1{--}2.3~keV, and estimate the typical impulsive event frequencies, timescales, amplitudes, and the distribution of amplitudes. We find that the impulsive events occur at a frequency of $\approx$25 events per minute with a typical lifetime of $\approx10$ minutes. They are characterized by a power law distribution with a slope $\alpha\leq2.0$. The typical amplitudes of these events lie in an energy range of $10^{21}${--}$10^{24}$~ergs, with a typical radiative loss of about $\approx10^3$ erg~cm$^{-2}$~s$^{-1}$ in the energy range of 1{--}2.3~keV. These results provide further constraints on the properties of sub-pixel impulsive events in maintaining the quiet solar corona.

\end{abstract}


\section{Introduction} \label{sec:intro}
The anomalous heating of the solar corona occurs not just in bright active regions, but also in the Quiet Sun (QS) regions. Hence, the route to understanding the temperature of the corona begins by understanding the heating in QS. The two predominant physical mechanisms responsible for heating the solar corona are through dissipation of Magnetohydrodynamic (MHD) waves via resonant absorption or phase mixing \citep[see, e.g.][]{alfven_1947_waveheating,osterbrock_1961_MHDWaveheating,AntolinShibata} or that of magnetic stresses via magnetic reconnection or Joule Heating \citep[see, e.g.][and also \cite{klimchuk_2006_coronalheating, parnell_2012coronalheatingreview}]{parker_1972_Bfieldbraiding, tzihong_1987_reconnection, Parker_1988_nanoflares,Hansteen2010_SWModelling}. However, it has been shown that both of these mechanisms give rise to the heating in an impulsive manner \citep{AntolinShibata,Klimchuk_2015_Corheating}.

Impulsive events are very well observed in all layers of the solar atmosphere over a range of spatio-temporal scales~\citep[see, for example][]{Benz_1998_heatingeventsinQS,benz2008}. Such impulsive events are even potentially consistent with dynamics of transition region lines like \ion{Si}{4}~\citep[][]{GupST_2018,TriNS_2021,Tri_2021}, and chromospheric lines like the \ion{C}{2}~\citep{Vishal_2021_CII} and \ion{Mg}{2}~\citep[][]{Vishal_2022_QSCH}. They also typically explain the dynamics and coupling in the chromosphere and transition region, with potentially explaining the formation of the solar wind~\citep{Hansteen2010_SWModelling}, and structures like switchbacks~\citep[see][for more details]{Vishal_2022_QSCH}. Thus, an impulsively heated scenario may potentially explain the heating of the QS corona.

If impulsive events are to explain the temperature of the solar corona, they must follow a power law distribution in energy of the form: $\frac{dN}{dW}\propto W^{-\alpha},$ with $\alpha > 2$ as conjectured by \cite{hudson1991solar}. However, there is a wide range of $\alpha$ reported in literature \citep[see, for e.g.][]{kuhar2018nustar,aschbook,Alipour2022BP}, depending on the time, region and energy band of observation \citep[see also, for e.g.][]{berghmans_1998,parnell_2000,Hannah_2008_powerlaw,berghmans2021extreme}. 

The QS corona is generally very diffuse. The impulsive events thought to explain the QS are expected to be of the order of, or smaller than $\approx700$~km, and lie in the sub-pixel regime~\citep{PSModel}.  Typically, the individual events responsible for generating the QS corona may remain unresolved, and would not show a clear signature of resolved events like microflares.  Hence, any statistic or model based on naive counting of occurrence of individual events in a typical QS light curve, for example, introduces a bias either towards the larger events, or an under-counting of multiple small events as one large event. These individual events, however, leave a collective imprint on the entire light curve in a statistical manner. These imprints have been shown to be statistically reflected in the intensity distribution of light curves from Active Regions~\citep[see, for example the analysis by][]{Vekstein_2009_distribution_AR,Terzo_2011_distribution_AR,Jess_2014_chromosphere,jess2019} and coronal loops seen in X-rays~\citep{Sakamoto_2009ApJ_distribution_ARLoops}. Clearly, while individual events may not be measured, their cumulative effect on the statistical properties of intensity light curves can be leveraged to understand these events. Thus, the existence of such small-scale events may only be inferred statistically.

Typically, a `statistically-realistic' simulation would be the one which reproduces some salient properties of the observations well. A statistical and impulsively heated mechanism may leave signatures in the distribution of intensity, the characteristic temporal features, or in the thermal structure of plasma~\citep[see, for e.g.][]{Sturrock_1990_ref,hudson1991solar,Sylwester_2019_ref,Rajhans_2021_ImpulsiveEvent}.~\cite{hudson1991solar}, for example, show that the relative interplay of frequency of occurrence of events and the time scale of the events reflects in the temporal power spectrum of the emergent light curves. One such empirical model was proposed by \cite{PSModel}, hereafter referred to as the \texttt{Pauluhn and Solanki Model}({\psm}). {\psm} is an empirical model based on two key observations: the log-normal distribution of intensities in corona when taken spatially or temporally \citep{ps_2001,Andretta_DelZanna_Lognormal}, and the power law distribution of energies of individual events \citep[see e.g.,][]{aschbook}. In brief, the {\psm} constructs light curves through a combination of multiple impulsive events enforcing a Markovian process to generate these light curves. The amplitudes of the impulsive events are sampled from a power law, while the distribution of generated intensity is theoretically shown to be log-normal by \cite{PSModel}. For details of \psm, see \citet{ps_2001,PSModel}.

The {\psm} has been applied successfully to light curves from various regions in the solar corona \citep[see, for example][]{PSModel,safari2,safari1,VDModel}. In a nutshell, the application of {\psm} essentially involves inferring the free parameters of the model for a given light curve (or a set of light curves). 

Parameter estimation may be performed using different methods. \cite{hudson1991solar} demonstrate the effect of high- and low frequency heating on the temporal power spectrum qualitatively. However, it is desirable to actually infer the individual free parameters of the models given the light curves or power spectra, and not just obtain a qualitative match. For this purpose, we use machine learning, which is a paradigm for constraining the free parameters of a non-linear model using data \citep{Goodfellow-et-al-2016}. \cite{VDModel} developed an exhaustively validated machine learning inversion model for the {\psm}. Simply put, the inversion model employs a Convolutional Neural Network \citep[CNN;][]{lecun2015deep} to infer the parameters of {\psm} for given observed light curves as input. \cite{VDModel} successfully demonstrated their inversion model on a large number of light curves from Atmospheric Imaging Assembly \citep[AIA;][]{AIA2012} on board the Solar Dynamics Observatory \citep[SDO;][]{SDO}, and infer the distribution of different parameters for more than $300,000$ QS light curves. This inversion scheme, which we shall henceforth call {\ipsm}~\footnote{Inversion code for PSM}, can infer the free parameters of {\psm}, as evidenced by the excellent match in the intensity histogram and the M\'orlet wavelet power spectrum \citep{torrence1998practical} between the simulations and observations. 

The AIA data used by \cite{VDModel} were in units of DN/s, and did not have absolute flux calibration. Hence, while the parameters may be estimated well, it is difficult to ascertain the energy range over which these events occur. To mitigate this primary drawback, we consider the full-disk integrated, flux calibrated data from the Solar X-ray Monitor (XSM) onboard Chandrayaan-2 mission \citep[]{Vadawale14,shanmugam20,Mithun_2021_XSMcalibration} of the Indian Space Research Organization (ISRO). We apply the {\ipsm} on two light curves obtained in three energy bands during the solar quiet time. Furthermore, we also generate better constraints on the bounds of {\psm} power-law input by applying a simple metric-based search on top of {\ipsm}. This leads us to a better estimation of the energetics of the impulsive events in QS.

The rest of the paper is structured as follows: in \S\ref{sec:data}, we describe the dataset used in this analysis. In \S\ref{sec:model} we describe the forward model, uncertainty characterization and the inversion scheme. In \S\ref{sec:results} we report the results of our inversion and various properties of our results, and finally we discuss the consequences of our results in \S\ref{sec:discuss}.

\section{Observations and Data}\label{sec:data}
For the present work we have used the observations recorded by XSM on-board Chandrayaan-2 mission. XSM observes the Sun as a star and provides measurement of X-ray spectra in the energy range of 1{--}15~keV with an energy resolution of $\approx$175 eV at 5.9~keV and a time cadence of one second. It has been demonstrated that XSM has the sensitivity to carry out spectral measurements even when the solar activity is well below A-class \citep[]{Mithun20}. Thus, it is possible to use XSM observations to obtain X-ray flux from the Sun during quiet phases.

We have selected XSM observations for two time periods (Oct 17{--}21, 2019 and Feb 14{--}21, 2020) when there were no active regions on the solar disk as confirmed from \url{https://solarmonitor.org}. In this work, we are interested in studying the contribution of the unresolved impulsive events to the quiet coronal light curves and not the well resolved events like microflares. Thus, by visual inspection of the X-ray light curves, we removed the microflare like events studied by \citealp{Vadawale21a} so that the selected observations form a true representation of quiescent solar corona, similar to \cite{Terzo_2011_distribution_AR}. This step inevitably gave rise to data gaps. However, since we are interested in a statistical study of the QS light curves, we have concatenated the light curves by ignoring the gaps and obtained a continuous time series.

For the selected duration, we generated effective area corrected and time-resolved X-ray spectra from the raw data using XSM Data Analysis Software (XSMDAS; \citealp{mithun21_soft}). Given the very low solar X-ray flux during these observations, the time bin size for spectra was chosen to be 2 minutes so that uncertainties on the flux due to counting statistics is typically less than 5\%.
\begin{figure}
    \centering
    \includegraphics[width=\linewidth]{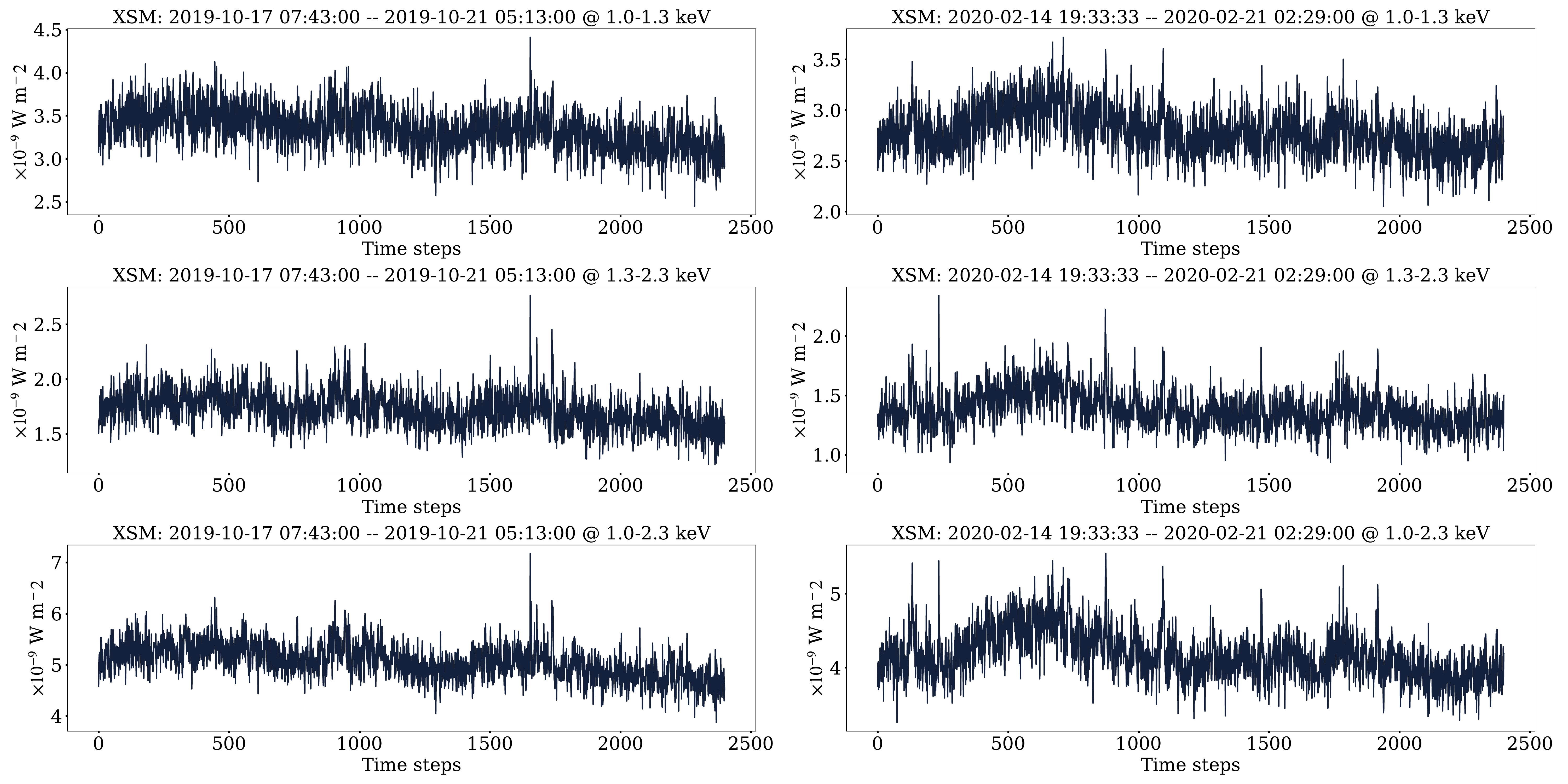}
    \caption{The XSM observed light curves considered in this work. The left(right) column is for 2019(2020) observations. Top row is the light curve for 1.0{--}1.3~keV, middle row is for 1.3{--}2.3~keV and the bottom row is for 1.0{--}2.3~keV.}
    \label{fig:observation}
\end{figure}

The X-ray flux light curve, $F(t)$, in the energy range $E_1$ to $E_2$ is then computed from the time-resolved spectra $S(E,t)$ as:
\begin{equation} \label{lc}
F(t) = \sum_{E={E_1}}^{{E_2}} \frac{S(E,t)~E}{A(E)}
\end{equation} 

\noindent where $A(E)$ is the on-axis effective area of the XSM \citep{Vadawale21a}. For both the observations, we generated light curves using eq.~\ref{lc} for the energy ranges of 1.0{--}1.3~keV, 1.3{--}2.3~keV, and 1.0{--}2.3~keV. The light curves so obtained are shown in Fig.~\ref{fig:observation}. Spectra above 2.3~keV are not considered as no appreciable flux is observed above that energy by XSM during QS observations.

\section{Modeling}\label{sec:model}
We seek to understand the properties of events that give rise to the X-ray light curves shown in Fig.~\ref{fig:observation}. For this purpose, we employ the {\psm}, which is based on the impulsive heating scenario. An impulsive event in the {\psm} is defined as an exponential rise of intensity to a peak value, and a subsequent exponential fall-off. This is determined by the total timescale $\tau$, which is the sum of the exponential rise ($\tau_r$) and the decay ($\tau_d$) timescale. These two time scales are related as $\tau_r = 0.75\tau_d$ following \cite{PSModel}. Note that while this relation is not exact, it is observationally motivated. \cite{Zhang_2015_timescalerelations} find a direct relation between rise and decay times, though this relation was a power law for the large flare cases considered. Also note that the constant $0.75$ has been put in ad-hoc by \cite{PSModel}. The peak value, i.e., the amplitude of each event, is sampled from a power-law distribution. This power law distribution has a slope $\alpha$, with a lower and upper energy limit $y_{min}$ and $y_{max}$, respectively. A multitude of impulsive events occurring at different times gives rise to the simulated light curves. The probability that an event will even occur at a time step is controlled by the  flaring frequency ($p_f$). Hence, the {\psm} has 5 free parameters, namely $p_f$, $\tau$, $\alpha$, $y_{min}$ and $y_{max}$, which fully determine a simulated light curve. This reduces our problem to the estimation of the above mentioned five parameters for XSM light curves shown in Fig.~\ref{fig:observation}. 

\begin{figure}
    \centering
    \includegraphics[width=\linewidth]{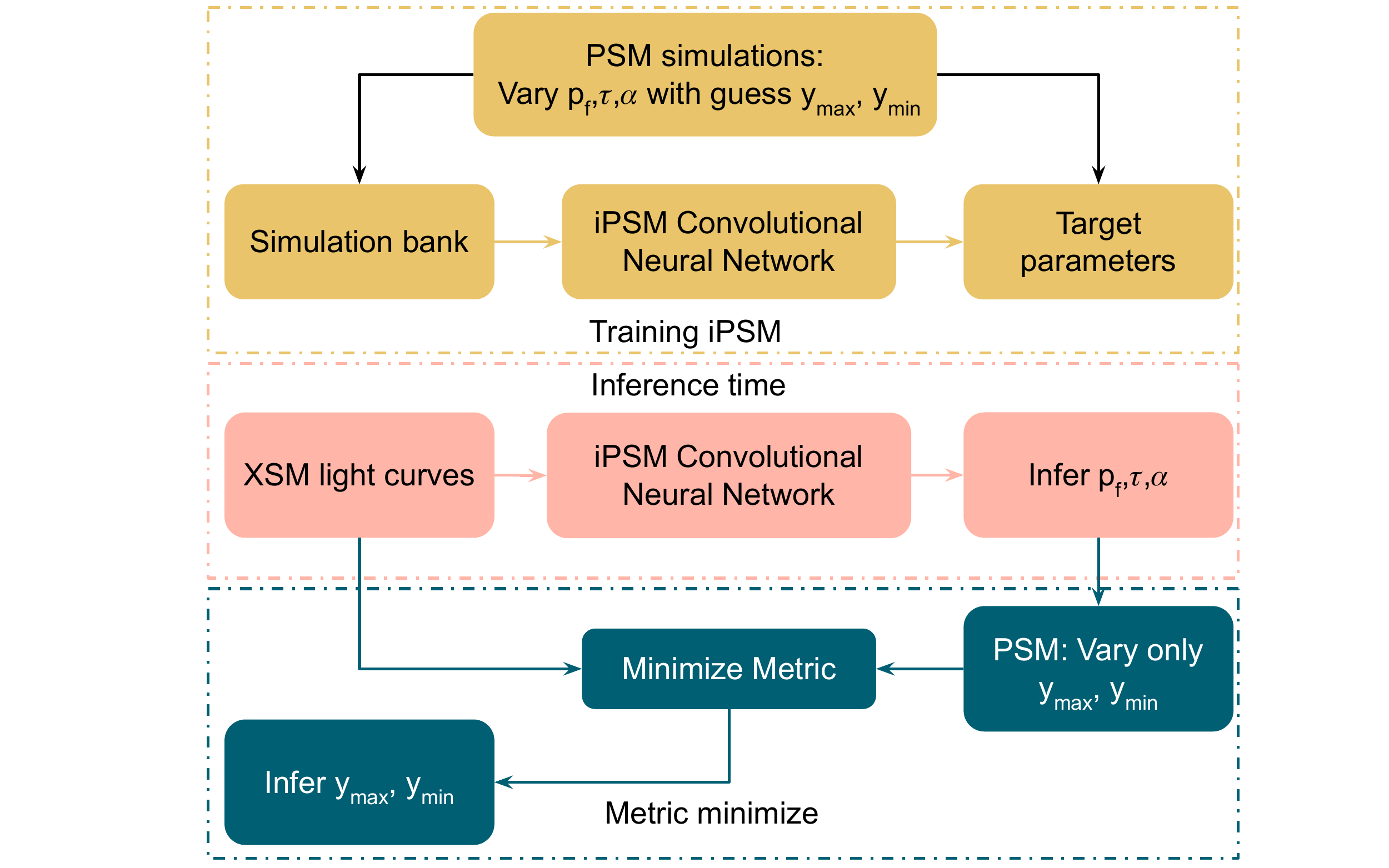}
    \caption{Flow chart detailing the various steps in our algorithm. First, the {\ipsm} is trained on simulations. Next, the trained model is used to infer $p_f$, $\tau$ and $\alpha$. Finally, the $y_{max}$ and $y_{min}$ are inferred by minimizing an error metric.}
    \label{fig:flowchart}
\end{figure}
The {\ipsm} model forms the core inference block of our work. The optimization for all the 5 parameters is inherently difficult to perform due to degeneracy in the parameter space. Hence, {\ipsm} performs inference of only three of the free parameters, while keeping $y_{min}$ and $y_{max}$ fixed. Hence, we breakdown the inversion scheme into two steps (see Fig.~\ref{fig:flowchart}), i.e., determining $p_f$, $\tau$ and $\alpha$ in the first step and $y_{max}$, $y_{min}$ in the second step by performing a fine search over the exact range of amplitude of the events. 

As shown in Fig.~\ref{fig:flowchart}, each step further consists of several parts. However, the first step requires we already have a reasonable estimate of $y_{max}$ and $y_{min}$. Hence, we first put reasonable bounds on range of values $y_{max}$ and $y_{min}$ can take by using prior observations, and fix an initial guess. Using this $y_{max}$ and $y_{min}$, we generate a bank of the {\psm} simulations sweeping across a range of $p_f$, $\tau$ and $\alpha$. We then use this simulation bank to train the {\ipsm} (the ``Training iPSM'' block in Fig.~\ref{fig:flowchart}), and learn the mapping from the simulated light curves to their corresponding parameters. Finally, we perform a forward pass of the XSM light curves through the trained model, and infer the corresponding values of $p_f$, $\tau$ and $\alpha$ (``Inference time'', pink colored section in Fig.~\ref{fig:flowchart}).

Using the inferred values of $p_f$, $\tau$ and $\alpha$ from step{--}1, in step{--}2, we generate another bank of simulations, this time sweeping on $y_{max}$ and $y_{min}$. Note that the range of $y_{max}$ and $y_{min}$ is within the bounds as described in step{--}1. Finally, by minimizing an appropriate metric, we perform a parameter sweep considering the XSM light curves to infer $y_{max}$ and $y_{min}$ (``Metric minimize'', sea green section in Fig.~\ref{fig:flowchart}). Thus, through a two-step process, we infer all the 5 free parameters of the {\psm}.
\subsection{Fixing $y_{max}$ and $y_{min}$}\label{sec:ymaxymin}

As described above, for generating the simulation bank for {\ipsm} in step{--}1, we need to fix $y_{max}$ and $y_{min}$. Furthermore, we need to define bounds of $y_{max}$ and $y_{min}$ over which the step{--}2 search is performed. To do so, we first fix the upper bound of $y_{max}$ and lower bound of $y_{min}$ approximately, and then fix the $y_{max}$ and $y_{min}$ values within this range for step{--}1. We first define the integrated energy per flare as:

\begin{equation}
    E = 4\pi\mathrm{R}_{1\mathrm{AU}}^{2}\cdot\tau\cdot\mathrm{F}_{\mathrm{median}}\cdot \mathrm{F}_{\mathrm{code}},
    \label{eq:F2E}
\end{equation}

where E is luminosity in a given energy band, $\mathrm{F}_{\mathrm{code}}$ is the amplitude of an event in code units, $\tau$ is the associated timescale (in seconds), $\mathrm{R}_{1\mathrm{AU}}$ the distance from Sun to Earth in meters, and $\mathrm{F}_{\mathrm{median}}$ the median intensity of the XSM light curve in W${\mathrm{m}^{-2}}$. Note that we divide the observed light curves by their median values during training and inference time, so the event amplitudes in code units would need to be multiplied by the same scaling to get the correct dimensional values. This conversion factor in Eq.~\ref{eq:F2E} helps us translate from the energy of an event in code units to real units. Since we want to generate bounds on $y_{max}$ and $y_{min}$, we fix the bounds for $\mathrm{F}_{\mathrm{code}}$, given other terms in Eq.~\ref{eq:F2E}.

Eq.~\ref{eq:F2E} has terms on the right hand side (except $\mathrm{F}_{\mathrm{code}}$) common for both the upper bound of $y_{max}$ and lower bound of $y_{min}$. Let us consider the median intensity in the 1{--}2.3~keV energy band, which is $\approx5\times10^{-9}$ $\mathrm{W}{\mathrm{m}^{-2}}$ (see Fig.~\ref{fig:observation}), 1 AU to be $\approx1.5\times10^{14}$~m, and a maximum time scale of $\approx720$~seconds. We obtain this timescale from the {\ipsm} inversions of light curves in the 211~{\AA} passband of QS, as obtained by \cite{VDModel}. The AIA 211~{\AA} passband corresponds to a temperature of $\log\, T \approx 6.2$, while the X-ray measurements typically lie in the range of  $\log\, T \approx6.2-6.8$~\citep{Santosh_XSMMicroflare}. Thus, we use the 211~{\AA} results as a proxy for the X-ray measurements here. Hence, an event with unit amplitude event (i.e  $\mathrm{F}_{\mathrm{code}}=1$) would correspond to an energy of $\approx10^{25}$~ergs.

First, we generate an upper bound for $y_{max}$. We note that in our dataset, all the microflares studied by \cite{Santosh_XSMMicroflare} have been removed. Hence, an individual event in any of our simulation cannot be larger than the smallest flare observed by \cite{Santosh_XSMMicroflare}. Since we are operating in particular energy band, we redo the energy distribution computation in \cite{Santosh_XSMMicroflare} for the energy band of 1{--}2.3 keV. The lowest energy thus inferred by \cite{Santosh_XSMMicroflare} for this energy band corresponds to $10^{24}$ ergs. This corresponds to `E' in Eq.~\ref{eq:F2E}. Thus, $\mathrm{F}_{\mathrm{code}}$ should be $<0.1$ for the maximum amplitude condition to be satisfied. Thus, we obtain an upper bound on $y_{max}$ -- i.e, $y_{max}<10^{-1}$. 

Having fixed the upper bound for $y_{max}$, we turn our attention to fixing the lower bound on the $y_{min}$. For this, we consider the energetics of fluctuations observed in soft-X ray light curves derived by \citealp{Katsukawa_yminlimits,Katsukawa_yminlimits_updated}. These authors found the energies of impulsive events of $\approx10^{20-22}$ ergs to be consistent with the distribution of fluctuations of soft-X ray light curves in active regions. \cite{Labonte_2007_KatsukawaCounter} however showed that the fluctuations in the light curves as obtained by \cite{Katsukawa_yminlimits} are consistent with noise. Thus, we take the lower limit of the possible energies, and set a lower bound on $y_{min}$ as $10^{20}$~ergs, where the lower limit basically corresponds to ``noise'' events. This value would correspond to $y_{min}>10^{-5}$ in code units, following Eq.~\ref{eq:F2E}. Thus, the event amplitudes may lie only between $10^{-5}$ and $10^{-1}$. Hence, these physical observations set the general bounds of the range of the expected energies of events.

We have now obtained the lower bound on $y_{min}$ and the upper bound on $y_{max}$. To fix the values of $y_{max}$ and $y_{min}$ in step{--}1, we prototype on a very limited combination of $y_{max}$ and $y_{min}$, generating one {\ipsm} model for each combination. Through visual inspection, we find $y_{max}$ and $y_{min}$ of $5\times10^{-3}$ and $10^{-4}$ (code units) to give us simulations which show a good match in the intensity distribution \& wavelet power spectrum with the XSM observation. Thus, we fix $y_{max}$ and $y_{min}$ to be $5\times10^{-3}$ and $10^{-4}$ for generating the bank of simulations for step{--}1 of our inversion. 

\subsection{Statistical uncertainty model}\label{sec:noise}
Using all the parameters discussed above, we generate simulated light curves that can be compared with the observed light curves from XSM. However, the QS is known to have weak emission in X-rays \cite[see, for example,][]{Brosius_1997_QSARcompare,Katsukawa_yminlimits,OdwDM_2010}, and is expected to have a non-negligible contribution of counting statistics. Therefore, for an objective comparison, the associated simulated light curves must be incorporated with these statistical uncertainties.

For this purpose, we estimate the statistical uncertainties on the light curves by propagating the Poisson error on the observed count for each light curve. Hence, for each observed light curve to be inverted, we know the signal and the associated uncertainty at each time step. 

To get a non-dimensional estimate of the uncertainty as a single number for the full light curve, we first calculate the uncertainty-to-signal ratio $\rm{r}_{t}$ at each time step for a given XSM light curve. This provides us a measure of the ``uncertainty fluctuation'' as a fraction of the signal. To extract the essential features of this fluctuation while keeping the workflow simple, we consider the mode of $\rm{r}_{t}$, denoted as \rm{r}. This is the estimate of uncertainty as a fraction of the observed signal for the full light curve. This is justified since the variation of the mean count rates during the observations have remained nearly constant. We obtain \rm{r} for each observed light curve. 


To incorporate this uncertainty into each simulated light curve,  we replace the intensity at each time step with a sample from a Gaussian distribution with a mean of the simulated intensity (from the {\psm}), and a standard deviation of \rm{r} times the intensity at that time step. This is justified as while the original photon counts follow a Poisson distribution, the flux values after integration over two minutes are expected to follow a Gaussian distribution. Since there are 6 light curves, we have 6 associated sets of simulations for each light curve. 

Finally, by taking care of all the steps explained above, we simulate the light curves. We generate simulated light curves for a duration of 2400$\times$120 =288000 seconds at a time cadence of 1 second, i.e., 1 code time step = 1 second. To minimize the effect of starting seed, we generate the simulations with extra 1600 seconds, and throw away the first and last 800 seconds. The simulations are then re-binned at 120 second cadence, giving rise to 2400 time points to match the observations.

We incorporate the statistical uncertainties to each simulated light curve, and normalize each by it's median value as a pre-processing step to the {\ipsm}. After parameter inference, we construct the best-matching simulations by multiplying the median normalized simulations with the median value of the corresponding observation. This gives us simulation light curves in the units of $W~m^{-2}$. Since we are scaling the intensities in the simulations, we also scale the corresponding $y_{max}$ and $y_{min}$ values, which determine the amplitude of these events, in the same way. 

For step{--}1, we generate the bank of simulations by varying $p_f$ between 3 to 57 events per minute, which translates to $p_f$ between 0.05 to 0.95 events per second in steps of 0.01 events per second. The time scale $\tau$ is varied between 1 second and 500 seconds in steps of 10 seconds, while $\alpha$ is varied between 1.5 and 3.0 in steps of 0.1. This parameter space is similar to \cite{VDModel}. We have however reduced the maximum value of $\tau$ (in seconds), since we expect X-ray observations to show much shorter time scales than EUV observations, as seen by \cite{VDModel}. With the step{--}1 simulation bank ready, we are ready to perform step{--}1 inference of {\psm} parameters.

\subsection{Inversion scheme: The {\ipsm}}\label{sec:inversion}
The {\ipsm} is a deep-learning model as defined in \cite{VDModel}. This is a 1-D CNN model which takes in the light curves, and maps them to their corresponding free parameter set -- $p_f$, $\tau$ and $\alpha$. We describe the network architecture briefly below,  while the details are exactly as described in \cite{VDModel}. In essence, {\ipsm} contains multiple convolution layers sandwiched with non-linear transformations called activation functions. This forms the basic ``learning ability'' of the {\ipsm}.

During training time, a cost function is minimized to train the trainable parameters (weights) of the network. The hyper-parameters, which must be fixed by hand, are the same as those used in \cite{VDModel}. However, note that since the simulations in this work need to be uncertainty-incorporated, we retrain the model from scratch for the new set of simulated light curves. We generate 6 inversion models in total corresponding to each light curve. 

The {\ipsm} also generates epistemic uncertainties associated with every inversion. This is accomplished by randomly switching off certain neurons during inference time~\citep{hinton2012improving}, and performing a Monte Carlo forward pass. The obtained uncertainty is a reflection of how much weight - parameter space has not been explored by the {\ipsm}~\citep{gal2016uncertainty}.

The simulated light curves for each model are split into training set (80\%) and testing set (20\%). The model is trained on the training set, while the evaluation on the testing set is used to mark the convergence of the model. Following \cite{VDModel}, we use the coefficient of determination ($R^2$) as a measure of goodness of fit of the model. Simply put, $R^2$ performs a point-wise comparison between two arrays. If the two arrays are perfectly correlated, $R^2 = 1$. The worser the correlation, lower is $R^2$. We compare the actual simulation parameters with those obtained by inversion from the {\ipsm}. All of our models show $R^2>0.98$ for $p_f$ and $\tau$, while the $R^2$ for $\alpha$ are more than $0.91$. This step is depicted graphically as the yellow and pink flow diagrams in Fig.~\ref{fig:flowchart}. Thus, we infer $p_f$, $\tau$ and $\alpha$ for each XSM light curve from step{--}1.
\subsection{Inversion scheme: Metric minimization}\label{sec:metric_minimize}
In step{--}1, we have inferred three parameters $p_f$, $\tau$ and $\alpha$ for fixed values of $y_{min}$ and $y_{max}$. In step{--}2, we fix these three parameters and generate a new set of light curves by sweeping $y_{max}$ and $y_{min}$. We sweep $y_{max}$ between $9\times10^{-4}$ and $5\times10^{-2}$ with 45 steps in log$_{10}$, and $y_{min}$ between $1\times10^{-5}$ and $5\times10^{-4}$ for 36 steps in log$_{10}$. On these light curves we incorporate the photon counting uncertainties as described in \S\ref{sec:noise}. These light curves serve as a bank from which we may perform an inexpensive, simple search to generate better constraints on $y_{min}$ and $y_{max}$. However, to do so, we need to define a metric which we may then minimize. Since our qualitative ``best fit'' is determined by a good match between the simulation and observation in terms of intensity distribution and power spectrum, we define a simple metric in Eq.~\ref{eq:metric} as:

\begin{equation}
    m = \mathrm{max} \left((\mathrm{CDF}_{\mathrm{O}}-\mathrm{CDF}_{\mathrm{S}})^2\right)  + \mathrm{max} (\left((\mathrm{P}_{\mathrm{O}}-\mathrm{P}_{\mathrm{S}})/\mathrm{P}_{\mathrm{O}}\right)^2).
    \label{eq:metric}
\end{equation}
Here the subscripts \rm{O} and \rm{S} correspond to observation and simulation respectively. The first term finds the maximum of absolute difference between the cumulative distribution function of the two light curves. The second term finds the maximum relative wavelet power mismatch between the two light curves. 

With this metric, we then perform a grid search, and find the combination which gives us the lowest possible metric value. The corresponding $y_{max}$ and $y_{min}$ are then taken up as the `inferred' final values.
\section{Analysis and Results}\label{sec:results}
\subsection{Light curve inversions}
\begin{figure}[htpb!]
    \centering
    \includegraphics[width=\textwidth]{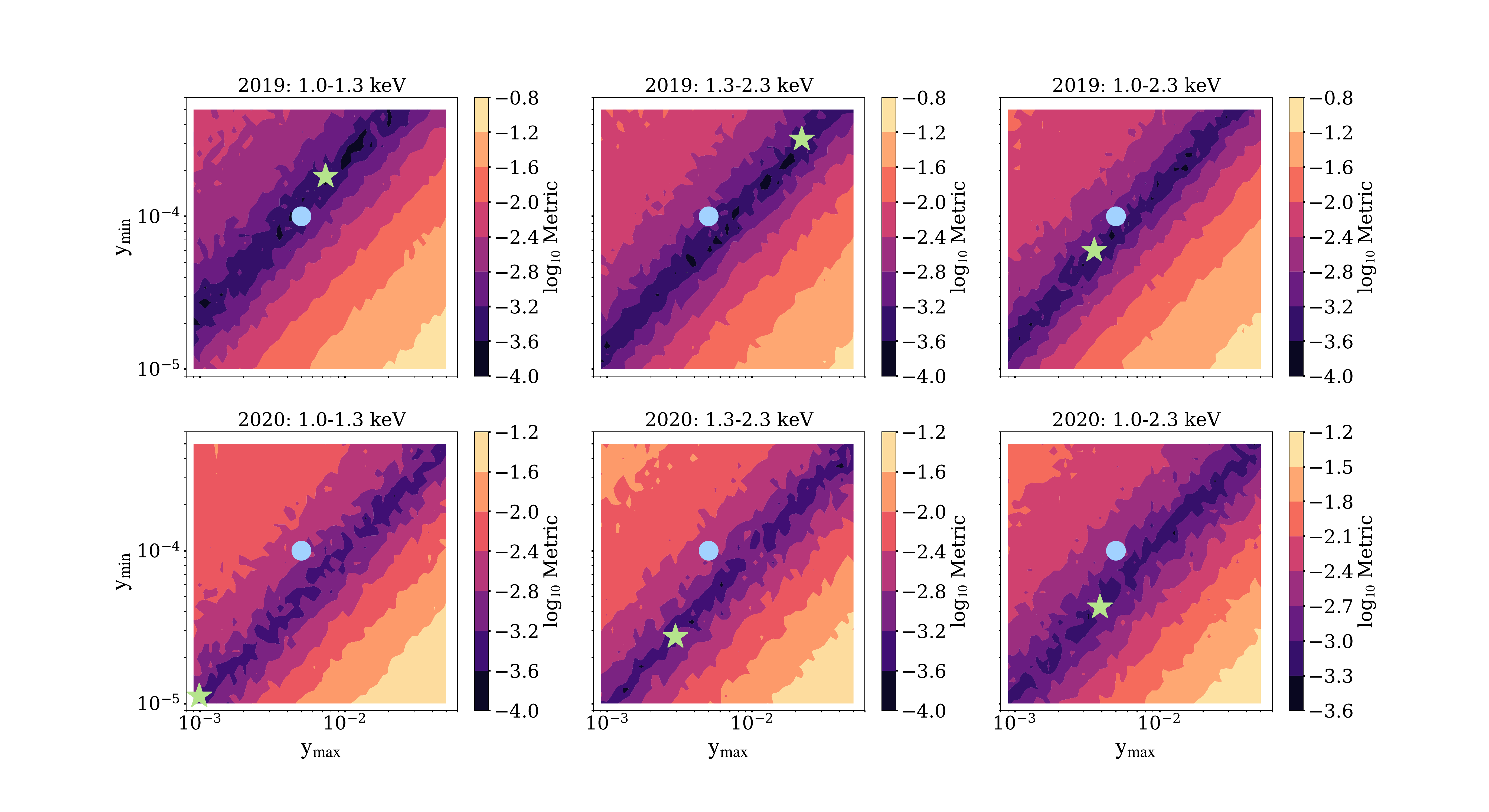}
    \caption{The variation of metric with $y_{max}$ (x-axis) and $y_{min}$(y-axis). The two parameters are presented here in code units (note the log scale), while the metric is presented in scale of $\log_{10}$. The blue circle shows the originally selected $y_{max}$ and $y_{min}$ (as used by the {\ipsm}), while the green star corresponds to be combination with lowest metric value.}
    \label{fig:ymaxminsweep}
\end{figure}

Applying our two step procedure described in \S\ref{sec:model}, we obtain the ``best fit'' parameters of the {\psm} simulations. In Fig.~\ref{fig:ymaxminsweep}, we present the metric surface from step{--}2 as a function of the swept range of $y_{max}$ and $y_{min}$, where the metric value is lower for darker color. Note that we have displayed the metric in log-scale. The blue circle represents the originally pre-fixed $y_{max}$ and $y_{min}$ for step{--}1, while the green star is the $y_{max}$ and $y_{min}$ solution inferred from step{--}2 parameter search. 

Fig.~\ref{fig:ymaxminsweep} reveals a number of salient features about our inferred solution(s). First, there is a whole diagonal of ``good'' solutions, showcasing the degeneracy between $y_{max}$ and $y_{min}$. Second, the pre-fixed $y_{max}$ and $y_{min}$ lie very close the diagonal ridge of good solutions, thereby also justifying our choice of initial guess for $y_{max}$ and $y_{min}$. Third, the final good solutions are sometimes quite close to the pre-fixed values, while sometimes they change by an order of magnitude. The final amplitudes, however, would depend on the median flux value. Therefore, the constraint is strongly performed for the ratio of $y_{max}$ and $y_{min}$. On the whole, a very strong global minimum is not clearly seen for constraining $y_{max}$ and $y_{min}$. However, the solutions as we shall see next give rise to a very good representations of the observed light curves.

We now present the inversion results for the two light curves obtained by integrating the signal between 1{--}2.3~keV energy band in Fig.~\ref{fig:2019_3}. The results for other light curves are presented in the Appendix~\ref{app:lcomp}.

In Fig.~\ref{fig:2019_3}, we show the light curves (panel a), intensity distributions (panel b), wavelet power spectrum (panel c) and cumulative distribution function (CDF; panel d). The orange represents the observation and the black represents the {\psm} forward model of the best fit parameters. Note that simulated light curve is uncertainty-incorporated. The uncertainty bands in the power spectrum correspond to 1-$\sigma$ standard deviation in time. The top four panels are for data recorded in 2019 and the bottom four are for that in 2020. Note again that a statistically accurate simulation must capture the intensity distribution well. Similarly, such a simulation must also capture the essential frequencies in the time series which have excess power. These are represented by the histogram (and CDF) and the wavelet power spectrum. The presence of peaks at similar frequencies in the power spectrum gives us the scales of importance, though we emphasize that the exact amount of power need not exactly match. The plots reveal a good correspondence between the observed and simulated light curve, both in matching the distribution, and wavelet power at different scales. Thus, the two-step inversion scheme with {\ipsm} successfully in captures the necessary information from the presented observations.

We summarize the inversion parameters for all the six light curves in Table.~\ref{tab:invertedparams}. We note that the flaring frequency $p_f$ ranges from $24-35$ events per minute and the time scale $\tau$ ranges from $\approx6-12$ minutes with maximum uncertainty of the order of a minute. For all the light curves, the inversion gives us power law slopes of $\leq2.0$. Finally, $y_{max}$ generally ranges from $7\times10^{-11}$--$1.26\times10^{-10}$~W~m$^{-2}$, while $y_{min}$ ranges from $6\times10^{-13}$--$2\times10^{-12}$~W~m$^{-2}$.
\begin{figure*}[ht!]
    \centering
    \includegraphics[width=\linewidth]{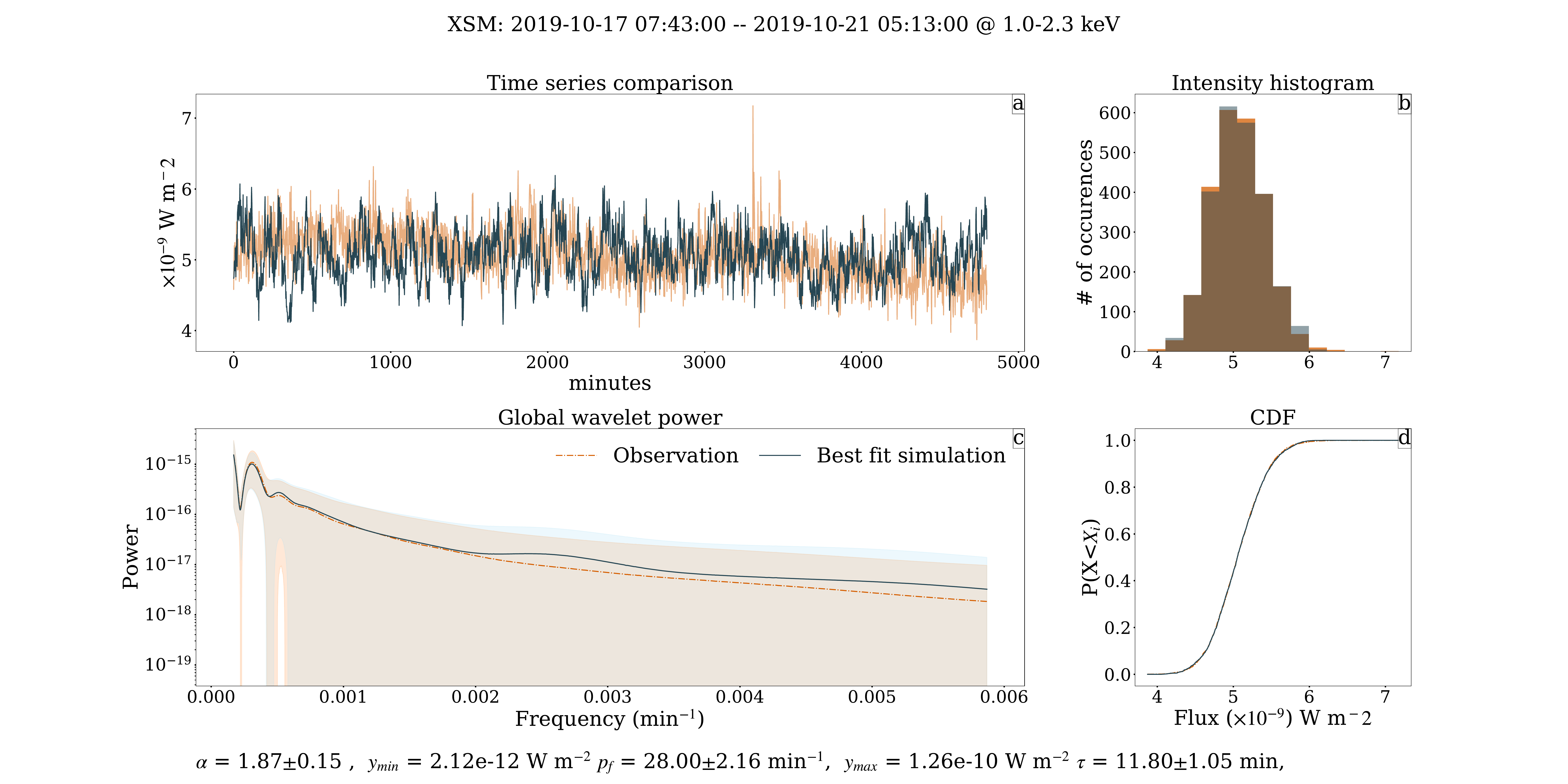}
    \rm{I}. 1-2.3 keV from October 2019
    \includegraphics[width=\linewidth]{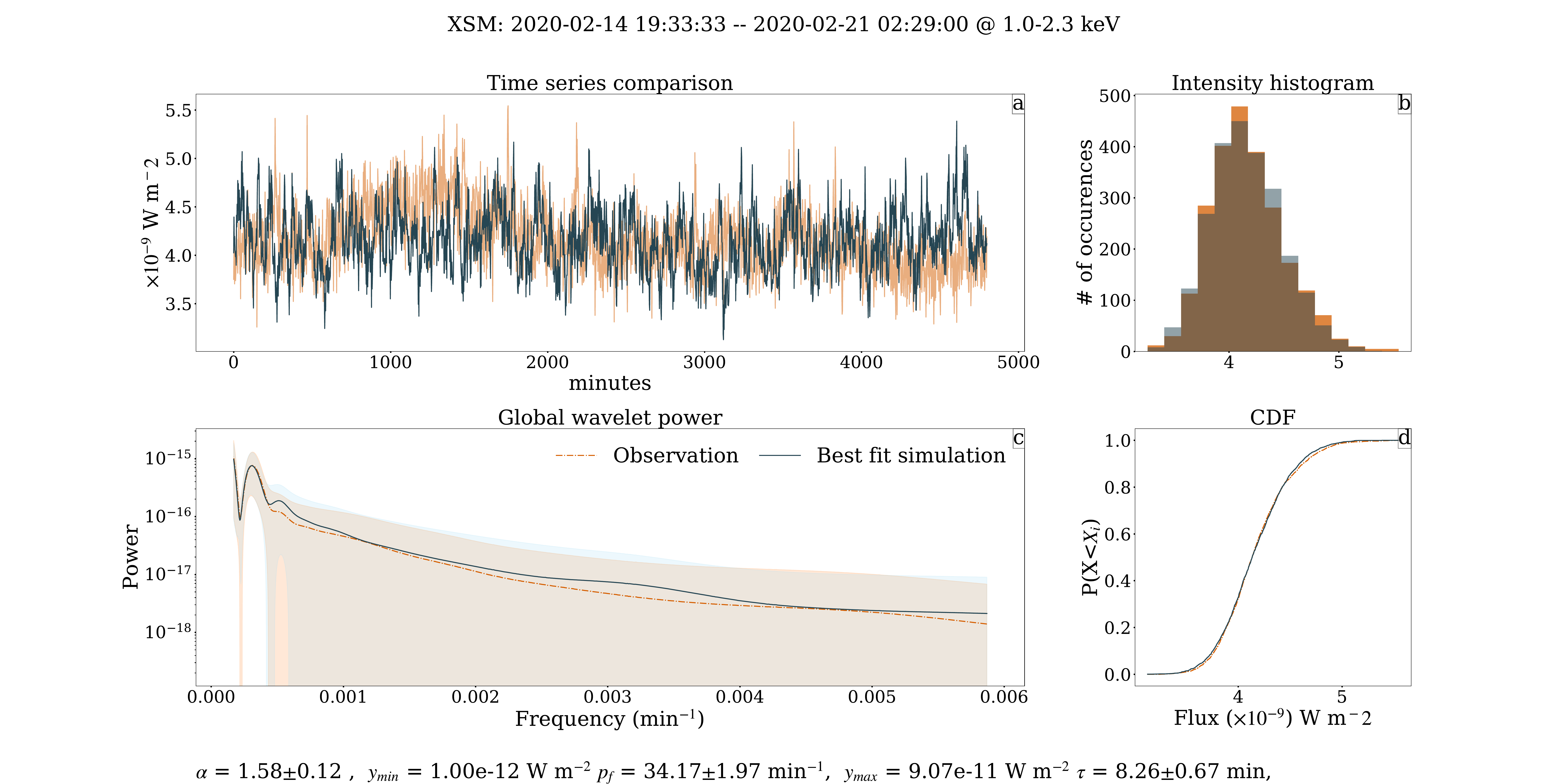}
    \rm{II}. 1-2.3 keV from February 2020
    \caption{Comparison of the observed light curve from XSM (orange), and the {\psm} forward model of best fit parameters inferred from our inversion code (black) in 1-2.3 keV energy band from 2019 (sub figure: \rm{I}) and 2020 (sub figure: \rm{II}). Each sub-figure has four panels depicting: Panel (a): Observed and simulated light curves; Panel (b): Distribution of observed and simulated light curve intensities; Panel (c): Global Morlét power for observation and simulations, with the uncertainties presented in orange and blue bands; Panel (d): Comparison of simulation and observation intensity CDF. The inset reports the inferred parameter set for the respective data.}
    \label{fig:2019_3}
\end{figure*}
\begin{table}[ht!]
    \centering
    \caption{Summary of the inferred parameters for the six light curves.}
    \label{tab:invertedparams}
    \begin{tabular}{c|cc|cc|cc}
        \hline
        \multirow{2}{*}{Parameter} & \multicolumn{2}{|c}{1.0 -- 1.3 keV} & \multicolumn{2}{|c}{1.0 -- 2.3  keV} & \multicolumn{2}{|c}{1.3 -- 2.3  keV}\\
        \cline{2-7}
         & 2019 & 2020 & 2019 & 2020 & 2019 & 2020 \\
        \hline 
        $p_f$ (events min$^{-1}$) & $27.89 \pm 1.67$ & $33.18 \pm 1.87$ & $28.00 \pm 2.16$ & $34.17 \pm 1.97$ & $25.42 \pm 1.57$ & $24.95 \pm 5.15$ \\ 

        $\tau$ (min) & $10.56 \pm 0.88$ & $9.12 \pm 0.73$ & $11.80 \pm 1.05$ & $8.26 \pm 0.67$ & $9.29 \pm 0.79$ & $6.56 \pm 0.56$ \\ 

        $\alpha$ & $2.00 \pm 0.12$ & $1.74 \pm 0.15$ & $1.87 \pm 0.15$ & $1.58 \pm 0.12$ & $1.94 \pm 0.13$ & $1.56 \pm 0.13$ \\ 

        $y_{max}$ (W m$^{-2}$) & $8.14\times10^{-11}$ & $7.30\times10^{-11}$ & $1.26\times10^{-10}$ & $9.07\times10^{-11}$ & $7.71\times10^{-11}$ & $6.86\times10^{-11}$ \\ 

        $y_{min}$ (W m$^{-2}$) & $2.03\times10^{-12}$ & $8.28\times10^{-13}$ & $2.12\times10^{-12}$ & $1.00\times10^{-12}$ & $1.12\times10^{-12}$ & $6.36\times10^{-13}$ \\ 
        \hline 
    \end{tabular}
\end{table}
\subsection{Energetics}
We now have a train of events giving rise to each of the observed light curves. Our goal is to study the energetics of these events. For this purpose, we first convert the obtained intensities into fluxes and energies following Eq.~\ref{eq:F2E}. Since we would be integrating only in a particular energy band, they would correspond to a ``lower bound" of energy. The energy estimates are better representatives of the energy content of these events if larger energy bands are considered. Hence we consider the energies in the widest 1{--}2.3~keV~passband. We find that our energies typically range between $10^{21}$--$2\times10^{23}$~ergs for this passband, with $\alpha$ shallower than $2.0$. These events will thus correspond to the nanoflare or even picoflare energy range. 

To understand the average radiative loss flux, we define the average amplitude of flare in a given time series ($A$) following \cite{PSModel} as :

\begin{equation} \label{eq:eng}
    \mathrm{A} := \left(\frac{1-\alpha}{2-\alpha}\right)\cdot\left(\frac{y_{max}^{2-\alpha}-y_{min}^{2-\alpha}}{y_{max}^{1-\alpha}-y_{min}^{1-\alpha}}\right)
\end{equation} 

The amplitude of the flare as defined in eq.~\ref{eq:eng} is in code units, which can be converted into real units of energies following Eq.~\ref{eq:F2E}. Inherently, we assume that the corresponding energy obtained is emitted isotropically by the Sun. To estimate the amount of energy emitted across the whole time series, we also need the frequency of occurrence of these events ($p_f$). Hence, for a given flaring frequency of $p_f$ (events per second), the amount of energy radiated  per unit time would be $p_f\cdot E$. Thus, the radiative flux loss from unit solar area (since we are performing full-disk integration) would be 

\begin{equation} \label{eqn:radflux}
    \mathrm{RL} := \frac{p_f4\pi\mathrm{R}_{1\mathrm{AU}}^2\cdot\tau}{A_{\odot,disk}}\mathrm{A}
\end{equation}

We find the radiative flux losses to be $\approx5\times10^3$ $\mathrm{erg}~\mathrm{cm}^{-2}~\mathrm{s}^{-1}$ in the 1-2.3 keV energy band, while they are $\approx3.5\times10^3$ and $\approx2\times10^3$~$\mathrm{erg}~\mathrm{cm}^{-2}~\mathrm{s}^{-1}$ in the 1{--}1.3 and 1.3{--}2.3~keV energy bands, with errorbars on each term. While the full set of results are presented in the Appendix in Table.~\refeq{tab:radloss}, the losses are typically of the order of $10^3$ $\mathrm{erg}~\mathrm{cm}^{-2}~\mathrm{s}^{-1}$.

\section{Discussion}\label{sec:discuss}
In this paper we study the QS heating and its energetics using XSM observations. To this end we use the empirical impulsive heating scheme of {\psm} as the ground truth. We have deployed a two step inversion scheme using the machine learning {\ipsm} coupled to a metric-based parameter search to infer the {\psm} parameters of our QS light curves. This inversion scheme let us infer the flaring frequency ($p_f$), the time scale ($\tau$), and the power law slope ($\alpha$), as well as the bounds of the power law ($y_{max}$, $y_{min}$) for any given light curve. By incorporating uncertainty model for the observed light curves, we perform the inference on full-disk integrated X-ray observations. The obtained results are summarized in Table~\ref{tab:invertedparams}.

We find that the flaring frequency is $\approx24-35$ events per minute. This flaring frequency is $10\times$ larger than those reported by \cite{VDModel} based on AIA observations, who found $p_f\approx2.5$ events per minute. These two results may be reconciled by noting that the average flaring amplitude (A) defined in Eq.\ref{eq:eng}, is $\approx10^{-3}$ in this study, while \cite{VDModel} find $A\approx10^{-2}$ in EUV. This shows that an approximately $10\times$ reduction in the flare amplitude has resulted in an approximately $10\times$ increase in flaring frequency ($p_f$). This is consistent with the finding of \cite{VDModel} that $p_f$ is found to reduce with increasing event amplitude. Thus, results for flaring frequency ($p_f$) obtained here for X-ray observations is consistent with their EUV counterparts shown in \cite{VDModel}, and strongly indicate the presence of an energy reservoir that may be depleted by large events occurring infrequently, or small events occurring more frequently. This flaring frequency translates to $\approx300$ events in a full-disk quiet coronal image in the 1-2.3 keV energy range, if an integration over $\approx11$ minutes is performed. However, note that we have selected extremely quiet times in this analysis, where there are no visible flares. Thus, the whole solar corona need not be at temperatures emitting strongly in the 1-2.3 keV energy range of X-rays. Furthermore, these events would have a typical amplitude of $10^{-13}${--}$10^{-11}$ W m$^{-2}$. This implies that even for regions emitting most strongly in X-rays in these quiet times, the events may not be detectable discretely upon incorporation of noise and statistical uncertainties. Thus, we expect only as a diffuse background to be seen with current instrumentation.

We further find that the event timescale ranges from $\approx6-11$ minutes. Based on EUV observations from AIA, \cite{VDModel} showed that the event time scales reduce with increasing temperature, i.e., from $\approx16$ minutes in 171~{\AA} ($\log\mathrm{T}\approx5.85$) to $\approx12$ minutes in 211~{\AA} ($\log\mathrm{T}\approx6.2$). Since the observations reported here are at higher temperature, the obtained results are consistent with those from EUV. However, note that these are only the mean values of a distribution of time scales as inferred by \cite{VDModel}.

We may also compare the properties of these unresolved events with those of resolved microflares.\cite{Sylwester_2019_ref}, for exmaple, studied microflares in the 1.2{--}15 keV energy range using data from SphinX \citep{Sylwester_2008_Sphinx}, with similar events studied by \cite{Vadawale21a}. They find the median temperatures of $\log\mathrm{T}\approx6.3$, while the time scales range from $\leq1$ minute to $\approx10$ minutes. Thus, the timescales we obtain are typically of the order of, or even slightly longer than those obtained by \cite{Sylwester_2019_ref} -- though we emphasize that timescales are consistent within the uncertainties.

Finally, we obtain $\alpha$ between $2.0$ and $1.56$ in this study, which are far flatter than those obtained by \cite{VDModel}, who find $\alpha\ge2$. However, note that the median $\alpha$ in \cite{VDModel} varies from 2.26 in 171~{\AA} to 2.07 in the 211~{\AA} passband. Consistent with this trend, we also find the $\alpha$ from X-rays to be smaller. Moreover, from Table.~\ref{tab:invertedparams}, we see that the $\alpha$ value reduces with increasing energy (from 1{--}1.3~keV to 1.3{--}2.3~keV). However, we note that the increase is only in the mean value, but within the errorbars they are consistent. On the whole, there appears to be a particular flattening of $\alpha$ with the increasing temperature of plasma. Thus, our results are consistent with those of \cite{VDModel} based on AIA observations. 

We emphasize, however, that the smaller $\alpha$ for higher temperatures is intriguing, i.e., for a given range of amplitudes, a larger $\alpha$ would have infrequent outlier intensities. However, a smaller $\alpha$, as inferred here, implies that these outlier events start to become the norm, implying that the typical amplitude of events is nearly constant. Moreover, \cite{Santosh_XSMMicroflare} find the power law slope to agree with 2 for the microflare observations using the same instrument and for the same time periods. This raises the question of a possible change of the underlying mechanism of heating from higher energies to lower energies, which reflects as differences in the power law exponent.

We note that the obtained energy of the impulsive events are typically in the nanoflare/picoflare regime and vary in the range of $10^{21}${--}$2\times10^{23}$~ergs. These values are typically of the scale of thermal energy as measured by \cite{Sylwester_2010_ref_AR} for typical solar quiet times, though we note that we report only a conversion of luminosity to energy, and not the thermal energy itself. Considering a very small range of energies the actual value of $\alpha$ may not even have a strong meaning. It may simply suggest that the events of $\approx10^{23}$ ergs are dominant over events with energy $\approx10^{21}$ ergs. Therefore, it is imperative to not consider just the parameter $\alpha$, but also consider the radiative flux in the events to get a better estimate.

Due to the flux-calibrated data of XSM, we can estimate the radiative energy loss from the quiet corona. We find the flux to be $\approx10^{3}~\mathrm{erg}~\mathrm{cm}^{-2}~\mathrm{s}^{-1}$ for the full energy range of 1{--}2.3 keV. This flux is two orders of magnitude lower than the radiative loss estimates in the quiet corona by \cite{Withbroe_Noyes_Energyflux}. Prima-facie, it suggests that such sub-pixel impulsive events may not have enough energy to maintain the quiet corona. However, we must note that the energy estimate presented here only provides a lower bound, since the energy is radiated away in many wavelengths. A better way would be to estimate the ``thermal energy'' content of the impulsive events, which is not possible in our case due to lack of spatial content, i.e., a length measure along the line of sight.

A caveat exists in the estimation of only the radiative energy in the given passbands of XSM. Similar to \cite{VDModel}, it is possible that some events may generate a response only in particular energy bands. Hence, the energetics of these events must be estimated by considering multiple energy bands in tandem while also introducing spectroscopic filling factors. This is a possible trajectory for the improvement of the {\psm} in the future. 

\begin{acknowledgements}
We acknowledge the use of data from the Solar X-ray Monitor (XSM) on board the Chandrayaan-2 mission of the Indian Space Research Organisation (ISRO), archived at the Indian Space Science Data Centre (ISSDC). XSM was developed by Physical Research Laboratory (PRL) with support from various ISRO centers. The authors thank Aveek Sarkar (PRL) for various discussions. The authors also thank the anonymous referee for providing numerous constructive comments and suggestions. This research is partly supported by the Max-Planck Partner Group on the Coupling and Dynamics of the Solar Atmosphere of MPS at IUCAA. 
\end{acknowledgements}

\software{Numpy~\citep{numpy_nature,numpy1,numpy2}, Scipy~\citep{scipy}, Matplotlib~\citep{matplotlib}, Multiprocessing~\citep{multiprocessing}, Tensorflow~\citep{abadi2016tensorflow}, Cython~\citep{behnel2011cython}, Jupyter~\citep{jupyter}.}

\section{Appendix}
\subsection{Light curve comparison for other energy bands}\label{app:lcomp}
\begin{figure*}[h!]
    \centering
    \includegraphics[width=\linewidth]{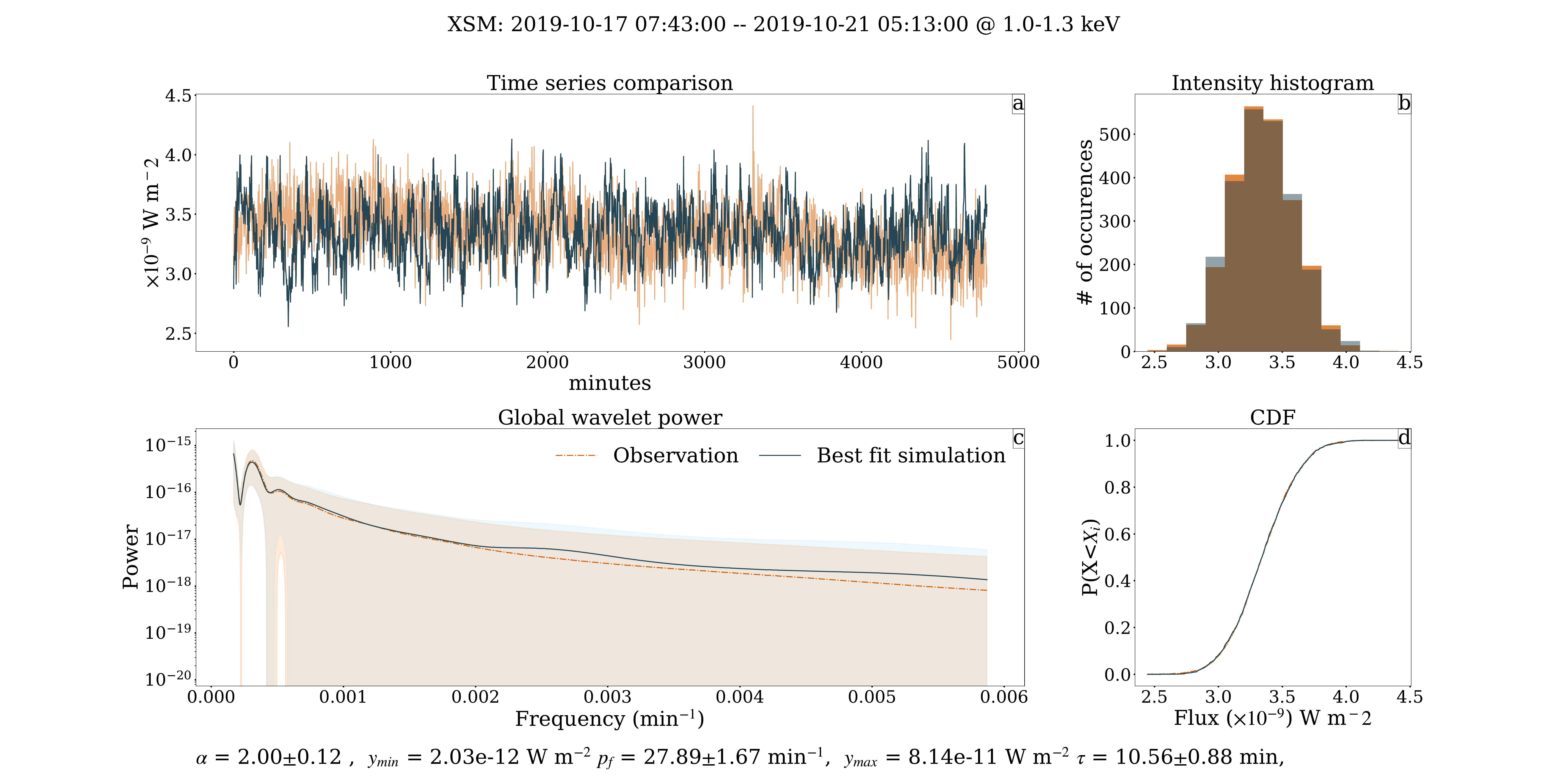}
    \includegraphics[width=\linewidth]{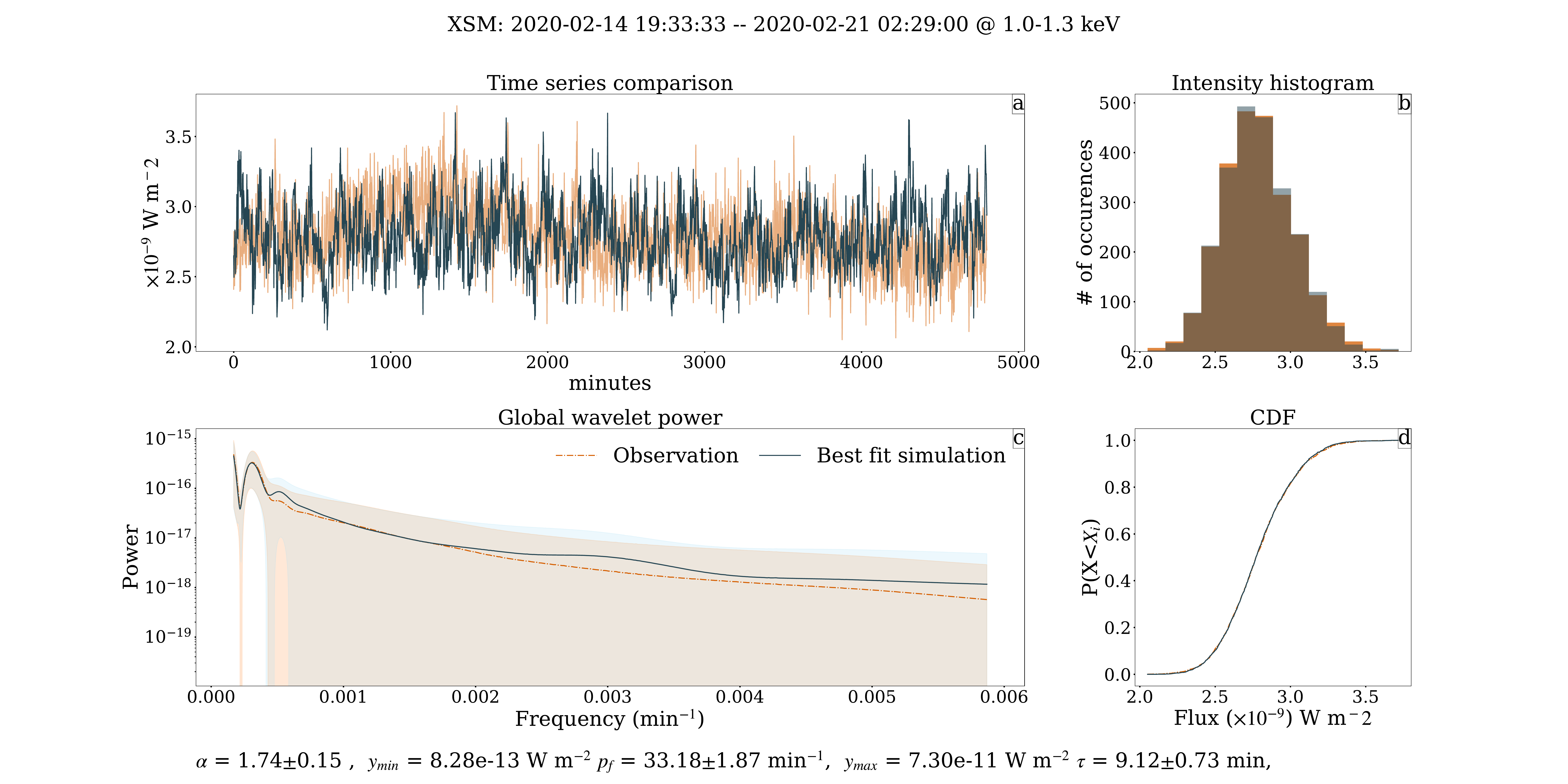}
    \caption{Same as Fig.~\ref{fig:2019_3}, but for 1.3{--}2.3 keV band.}
    \label{fig:2019_2}
\end{figure*}

\begin{figure*}[h!]
    \centering
    \includegraphics[width=\linewidth]{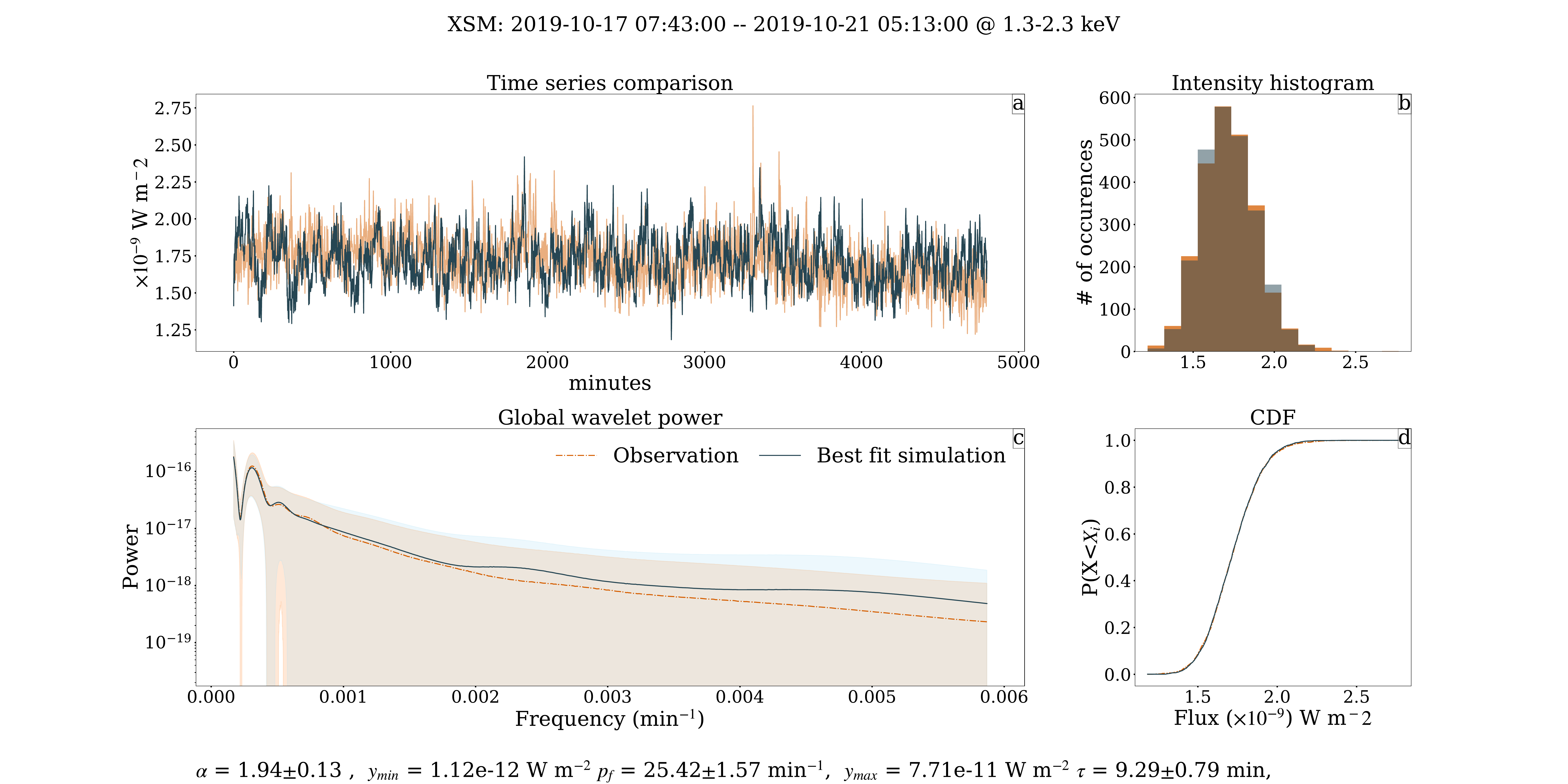}
    \includegraphics[width=\linewidth]{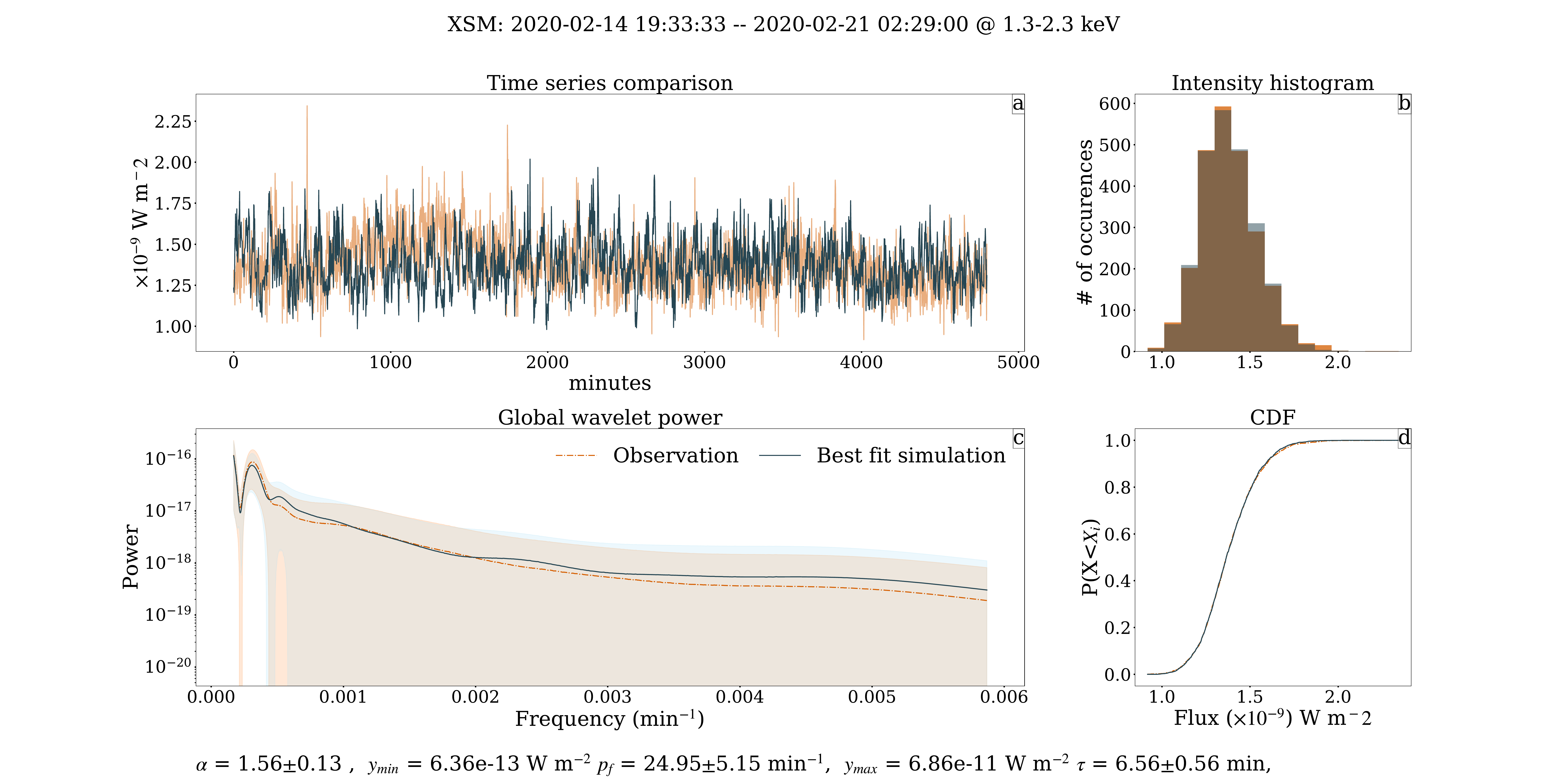}
    \caption{Same as Fig.~\ref{fig:2019_3}, but for 1.3{--}2.3 keV band for the segment from year 2020.}
    \label{fig:2020_2}
\end{figure*}

In this section, we present the light curve comparison for all the energy bands. The presentation follows same convention as Fig.~\ref{fig:2019_3}. These are presented as Fig.~\ref{fig:2019_2} and Fig.~\ref{fig:2020_2}.
\subsection{Distribution of events and their energetics}
\begin{figure}[h!]
    \plotone{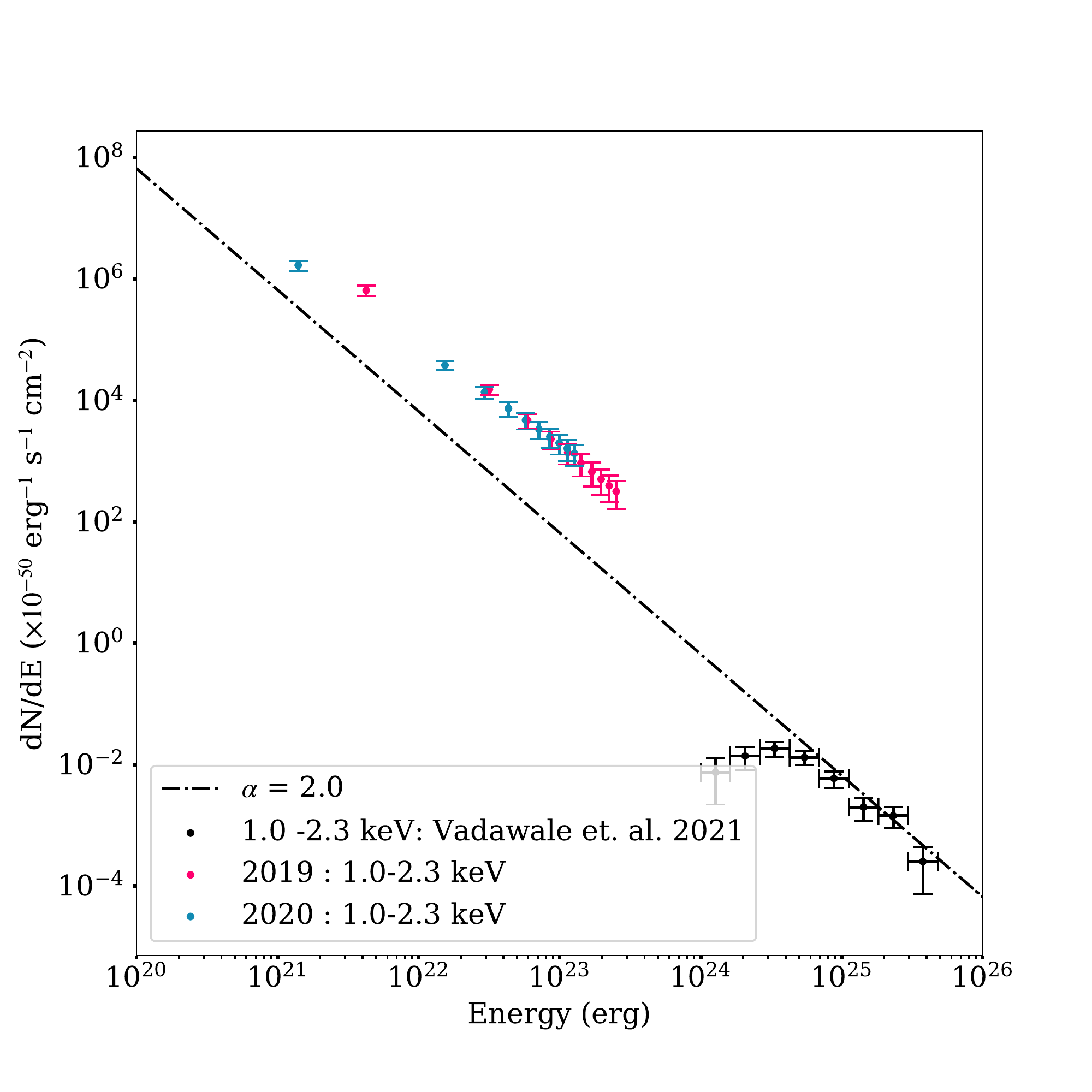}
    \caption{Frequency distribution of impulsive events inferred from the observations in 2019 year (pink) and 2020 (cyan). The scatter is frequency distribution of event energies from the model. The errorbars on inferred parameters are obtained by propagating the Monte Carlo uncertainties. The black scatter shows the inferred frequency distribution from \cite{Santosh_XSMMicroflare}, while the black dot-dashed line corresponds to $\alpha=2.0$.}
    \label{fig:power_law}
\end{figure}
\begin{deluxetable*}{ccc}[h!]
    \tablecaption{Radiative losses in $\mathrm{erg}~\mathrm{cm}^{-2}~\mathrm{s}^{-1}$ for the 3 energy passbands and two years. \label{tab:radloss}}
    \tablewidth{0pt}
    \tablehead{\colhead{Energy band (keV)} & \colhead{2019}& \colhead{2020}}
    \startdata
    1.0-1.3 & $4.18 \pm 0.65$ $\times10^3$ & $3.02 \pm 0.74$ $\times10^3$ \\ 
    1.0-2.3 & $6.29 \pm 1.37$ $\times10^3$ & $4.39 \pm 0.92$ $\times10^3$ \\ 
    1.3-2.3 & $2.26 \pm 0.43$ $\times10^3$ & $1.82 \pm 0.57$ $\times10^3$ \\ 
    \enddata
\end{deluxetable*}

In Fig.~\ref{fig:power_law}, we compare the distribution of the events from this work with those measured by \cite{Santosh_XSMMicroflare}, while in Table.~\ref{tab:radloss}, we report the radiative losses in all the energy bands for both the years. The errorbars are obtained by propagating the Monte Carlo uncertainties in the inverted parameters across the different equations.

Since we have an exact functional form for the event distribution, we sample $10$ points between $\mathrm{y_{min}}$ and $\mathrm{y_{max}}$, and plot our results for the 1{--}2.3 keV passband in Fig.~\ref{fig:power_law}. The pink color shows the data for 2019 and cyan color is for 2020, while the black color corresponds to the data by \cite{Santosh_XSMMicroflare}. The over-plotted dot-dashed back line represents a power law with $\alpha=2.0$.

\bibliography{references}{}
\bibliographystyle{aasjournal}



\end{document}